\documentclass[12pt]{article}
\usepackage{epsfig,amsmath,amssymb,mathrsfs}

  {\end{list}}%

\makeatletter
\renewcommand{\section}{\setcounter{equation}{0}\@startsection
 {section}%
 {1}%
 {0pt}%
 {-1\baselineskip}%
 {0.4\baselineskip}%
 {\bfseries\large}}%
\renewcommand{\subsection}{\@startsection
 {subsection}%
 {2}%
 {0pt}%
 {-0.75\baselineskip}%
 {0.2\baselineskip}%
 {\bfseries}}%
\renewcommand{\subsubsection}{\@startsection
 {subsubsection}%
 {3}%
 {0pt}%
 {-0.5\baselineskip}%
 {0.1\baselineskip}%
 {\sc}}%
\makeatother

\DeclareMathAlphabet{\mathpzc}{OT1}{pzc}{m}{it}

\def\msC{{\mathscr C}}

\def\calY{{\mathcal Y}}
\def\funY{{\mathpzc Y}}

\def\g5{\gamma_{5}}

\def\tpsi{\widetilde \psi}
\def\tphi{\widetilde \phi}
\def\tPsi{\widetilde \Psi}
\def\tPhi{\widetilde \Phi}
\def\tlambda{\widetilde \lambda}
\def\tLambda{\widetilde \Lambda}
\def\tSigma{\widetilde \Sigma}
\def\idx{\int\!\! d^4\!x}

\newcommand{\bea}{\begin{eqnarray}}
\newcommand{\eea}{\end{eqnarray}}
\newcommand{\beann}{\begin{eqnarray*}}
\newcommand{\eeann}{\end{eqnarray*}}
\newcommand{\ba}{\begin{array}}
\newcommand{\ea}{\end{array}}

 \def\g {\gamma}
\begin{document}
 \begin{titlepage}
\rightline{FTI/UCM 133-2010}
\vglue 33pt

\begin{center}

{\Large \bf Yukawa terms in noncommutative SO(10) and $\text{E}_6$  GUTs}\\
\vskip 1.0true cm
{\rm C. P. Mart\'{\i}n}\footnote{E-mail: carmelo@elbereth.fis.ucm.es}
\vskip 0.1 true cm
{\it Departamento de F\'{\i}sica Te\'orica I,
Facultad de Ciencias F\'{\i}sicas\\
Universidad Complutense de Madrid,
 28040 Madrid, Spain}\\
\vskip 0.85 true cm
{\leftskip=50pt \rightskip=50pt \noindent
We propose a method for constructing Yukawa terms for noncommutative SO(10) and
$\text{E}_6$ GUTs, when these GUTs are formulated within the enveloping-algebra formalism. The most general noncommutative Yukawa term that we propose contains, at
first order in $\theta^{\mu\nu}$, the most general BRS invariant Yukawa contribution  whose only dimensionful parameter is the noncommutativity parameter. This  noncommutative
Yukawa interaction is thus renormalisable at first order in  $\theta^{\mu\nu}$.
\par}
\end{center}

\vspace{9pt}
\vfill
\end{titlepage}

\section{Introduction}
The SO(10) and $\text{E}_{6}$ GUTs, which were introduced~\cite{gso10,
  Fritzsch:1974nn, Gursey:1975ki} in the mid 1970's,   are
the most popular GUTs in four dimensional space-time. They incorporate
right-handed neutrinos in the fermionic multiplets and realise the idea
of family unification --each Standard Model family snugly fits into an
irreducible multiplet, in addition to gauge coupling unification. These
theories can be made supersymmetric to achieve gauge coupling unification
after crossing the desert~\cite{Raby:2008gh,Senjanovic:2006zz}, but, may also --at least in the SO(10) case--
lead to nonsupersymmetric unification, if intermediate symmetry breaking scales (oases are thus
created in the desert) are introduced between the  electroweak scale and the
GUT scale~\cite{Lee:1994vp, Senjanovic:2006zz}. In view of all the results
obtained so far, and reviewed in~\cite{Raby:2008gh,Senjanovic:2006zz}, that GUTs may be relevant in the understanding of the data  which will
come out of the LHC is a thought that one cannot be rid of easily. A thought
that is also  prompted by the fact that SO(10) and $\text{E}_{6}$ GUTs arise naturally F-theory~\cite{Heckman:2009bi}.

More than a decade~\cite{Doplicher:1994tu, Seiberg:1999vs} has gone by since it became clear that field theories on
noncommutative space-time --which are named noncommutative field theories-- are to be considered in earnest. The formulation of
noncommutative gauge theories which are deformations of ordinary theories with
simple gauge groups in arbitrary representations demanded the introduction of
the enveloping-algebra formalism~\cite{Madore:2000en, Jurco:2000ja,
  Jurco:2001rq}  --a formalism which may find stringy accommodation in
F-theory~\cite{Cecotti:2009zf}.
The main feature of this formalism --see
ref.~\cite{Blaschke:2010kw}, for a review-- is that both
noncommutative gauge fields and infinitesimal noncommutative gauge
transformations take values on the universal enveloping algebra of the
corresponding Lie algebra, and are functions of the ordinary gauge fields;
these functions defining the corresponding Seiberg-Witten maps. The formulation of a
noncommutative generalisation --called the Noncommutative Standard Model-- of
the Standard Model demands the use of the enveloping-algebra formalism, if no
new particles are introduced --for noncommutative generalisations of the Standard Model outside the
enveloping-algebra formalism see refs.~\cite{Chaichian:2001py, Khoze:2004zc, Arai:2006ya}.
The Noncommutative Standard Model was put
forward in ref.~\cite{Calmet:2001na}, and a fair amount of
phenomenological consequences --which might be tested against the data from the LHC-- have been
drawn from it: refs.~\cite{Melic:2005su, Alboteanu:2006hh, Buric:2007qx,
  Tamarit:2008vy, Haghighat:2010up},
to quote only a few --the reader may wish to find further information in
ref.~\cite{Trampetic:2009vy}.
Renormalisability
~\cite{Buric:2005xe, Buric:2006wm, Buric:2007ix, Martin:2009sg, Tamarit:2009iy},
anomaly freedom~\cite{Martin:2002nr, Brandt:2003fx} and  existence of classical solutions~\cite{Martin:2005vr, Martin:2006px, Stern:2008wi} are other issues
which have been studied for noncommutative gauge theories formulated within the
enveloping-algebra formalism.

The general procedure to construct the noncommutative counterpart of the
ordinary SO(10) GUT within the enveloping algebra-formalism was laid down
in ref.~\cite{Aschieri:2002mc} --see also ref.~\cite{Bonora:2000td}. However,
the relevance in its phenomenological applications --footprints of a
noncommutative space-time may be found at the LHC-- of the Yukawa and Higgs
sectors of this theory demands that a detailed analysis and construction
of these sectors be carried out. At this point, we would like to stress that,
against all odds, theories which contain the fermionic and gauge
sectors --but have no Higgses-- of the noncommutative SO(10)
and $\text{E}_6$ GUTs are one-loop
renormalisable at first order in the noncommutativity parameter
--see ref.~\cite{Martin:2009vg}. So, it is a pressing issue to carry out a detailed construction of the first-order-in-$\theta$ Yukawa and Higgs sectors of these
theories, if the renormalisability properties of phenomenological relevant
noncommutative GUTs are to be studied. In this paper, we shall remedy this state of
affairs and propose a new strategy to construct the  noncommutative counterparts
of the ordinary SO(10) and$ \text{E}_6$ Yukawa terms that are renormalisable at
first order in the noncommutativity parameter. The ideas introduced here will be
certainly of help in the construction of the Higgs potential of noncommutative
SO(10) and $\text{E}_6$ GUTs, but, its construction will not be tackled here,
since it is very involved and surely deserves to be carried out separately.

The layout of this paper is as follows. In Section 2, we put forward our
procedure to construct noncommutative Yukawa terms for SO(10) and $\text{E}_6$
GUTs. In Section 3, we work out our noncommutative Yukawa terms at first order in the noncommutativity parameter taking into account
the symmetry properties, under the exchange of the fermionic multiplets,  of the
invariant tensor that occur in the ordinary Yukawa terms. Section 4 is devoted
to the discussion of redundant Yukawa terms. In Section 5, we state our  conclusions.

\section{Noncommutative Yukawa Terms for SO(10) and $\text{E}_6$}

In ordinary SO(10) and $\text{E}_6$ GUTs the fermionic degrees of freedom
are given by three fermionic field multiplets $\psi_{\alpha A f}$
--$\!f=1,2,3$, labels the three fermionic families of the GUT. For each $``A"$
and $``f",$ $\psi_{\alpha A f}$, $\alpha=1$ and $2$, denote, respectively,
the components of a left-handed Weyl spinor --here, we follow the conventions of
ref.~\cite{Dreiner:2008tw}; whereas, for each $``\alpha"$ and $``f",$ the index $``A"$
labels the components of the fermionic multiplet carrying certain --the 16, for
SO(10), and the 27, for $\text{E}_6$--  irreducible representations of the GUT gauge group.
The ordinary BRS transformations  of $\psi_{\alpha A f}$ are defined as follows:
\begin{equation}
s\psi_{\alpha A f}=i\lambda^{(\psi)}_{AB}\,\psi_{\alpha B f},\quad
s\lambda^{(\psi)}_{AB}=i\lambda^{(\psi)}_{AC}\,\lambda^{(\psi)}_{CB},\quad
\lambda^{(\psi)}_{AB}=\lambda^a\,\Sigma^a_{AB},
\label{brspsi}
\end{equation}
where $\Sigma^a_{AB}$ stands for a generic generator of the gauge group in the
representation furnished by the fermionic multiplet of each family.
We shall denote by $\phi_i$ the components of a generic Higgs multiplet
which  couples in the Yukawa terms to the fermions of our theory. We shall
assume that this multiplet  carries an irreducible representation of the GUT
gauge group. The BRS transformation of  $\phi_i$  is given by
\begin{equation}
s\phi_{i}=i\lambda^{(\phi)}_{ij}\,\phi_{j},\quad
s\lambda^{(\phi)}_{ij}=i\lambda^{(\phi)}_{ik}\,\lambda^{(\phi)}_{kj},\quad \lambda^{(\phi)}_{ij}=\lambda^a\,M^a_{ij},
\label{brsphi}
\end{equation}
where $M^a_{ij}$ denotes a generic generator of the GUT gauge group in the
irreducible representation supplied by the Higgs multiplet. As is well known,
for SO(10), $\phi_i$ will transform under either the 10, or the 120 or the
$\overline{126}$, whereas, the $27$, the $351'$ and  the
$351$
are the representations that may carry the Higgs multiplets in a Yukawa
term of the  $\text{E}_6$ GUT.

The ordinary Yukawa, $\calY^{\text{(ord)}}$, term for the gauge groups S0(10)
and  $\text{E}_6$ reads
\begin{equation}
\calY^{\text{(ord)}}=\idx\;\funY_{ff'}\;\msC_{AiB}\;\tpsi^{\alpha}_{A  f}\;\psi_{\alpha B f'}
\;\phi_{i},
\label{yukord}
\end{equation}
where $\funY_{ff'}$ denotes the Yukawa couplings and $\msC_{AiB}$ is
a group invariant three-index tensor, ie,
\begin{equation}
\tSigma^a_{AC}\,\msC_{CiB}\,+\,\msC_{AjB}
M^a_{ji}\,+\,\msC_{AjC}\,\Sigma^a_{CB}=0,
\label{invareq}
\end{equation}
where $\tSigma^a_{AC}\equiv\Sigma^a_{CA}$.
For later convenience, we have expressed
$\calY^{\text{(ord)}}$ in terms of
the $``A"$ component of the transpose of the fermionic multiplet $\psi^{\alpha}_{f}$:
$\tpsi^{\alpha}_{ f}=(\psi^{\alpha}_{f})^{\top}$ --of course,
$\tpsi^{\alpha}_{A f}=\psi^{\alpha}_{Af}$. The ordinary gauge transformations
act on $\tpsi^{\alpha}_{ f}$ on the right by means of the transpose matrix.
Hence the BRS variation of  $\tpsi^{\alpha}_{A f}$ reads
\begin{equation}
s\tpsi_{\alpha A f}=i\tpsi_{\alpha B f}\tlambda^{(\psi)}_{BA},\quad
s\tlambda^{(\psi)}_{BA}=-i\tlambda^{(\psi)}_{BC}\,\tlambda^{(\psi)}_{CA},\quad
\tlambda^{(\psi)}_{BA}=\lambda^a\,\tSigma^a_{BA},\quad\tSigma^a=(\Sigma^a)^{\top}.
\label{brstipsi}
\end{equation}

Let us now introduce  the following fields: $\phi_{AB}$, $\tpsi^{\alpha}_{iBf}$
and $\psi_{\alpha Aif'}$, which are defined as follows
\begin{equation}
\phi_{AB}=\msC_{AiB}\,\phi_{i},\quad\tpsi^{\alpha}_{iBf}=\tpsi^{\alpha}_{Af}\;
\msC_{AiB},\quad \psi_{\alpha Aif'}=\msC_{AiB}\;\psi_{\alpha Bif'}.
\label{reducifields}
\end{equation}
To construct noncommutative versions of $\calY^{\text{(ord)}}$ in
eq.~(\ref{yukord}), we shall find it  useful to have
$\calY^{\text{(ord)}}$ expressed in terms of the fields $\phi_{AB}$,
$\tpsi^{\alpha}_{iBf}$  and $\psi_{\alpha Aif'}$:
\begin{equation}
\begin{array}{l}
{\calY_1^{\text{(ord)}}\equiv \calY^{\text{(ord)}}=
\idx\;\funY_{ff'}\;\tpsi^{\alpha}_{ A  f}\;\phi_{AB}\;\psi_{\alpha B f'},}\\[4pt]
{\calY_2^{\text{(ord)}}\equiv \calY^{\text{(ord)}}=
\idx\;\funY_{ff'}\;\tphi_i\;\tpsi^{\alpha}_{i B  f}\;\psi_{\alpha B f'},}\\[4pt]
{\calY_3^{\text{(ord)}}\equiv \calY^{\text{(ord)}}=
\idx\;\funY_{ff'}\;\tpsi^{\alpha}_{A  f}\;\psi_{\alpha A i f'}\;\phi_i},\\
\end{array}
\label{yuk123}
\end{equation}
where, for later convenience, we have introduced $\tphi_i$, which is the
$``i"$component of the transpose of the Higgs multiplet: $\tphi=(\phi)^{\top}$.
 The fields $\phi_{AB}$, $\tpsi^{\alpha}_{iBf}$
and $\psi_{\alpha Aif'}$ do not carry irreducible representations of the GUT
gauge group, but they carry the very same number of physical degrees of freedom
as do $\phi_{i}$, $\tpsi^{\alpha}_{Bf}$ and $\psi_{\alpha A f'}$,
respectively. The BRS transformations of  $\phi_{AB}$, $\tpsi^{\alpha}_{iBf}$
and $\psi_{\alpha Aif'}$ are
\begin{equation}
\begin{array}{l}
{s\phi_{AB}=-i\,\tlambda^{(\psi)}_{AC}\,\phi_{CB}\,-\,i\,\phi_{AC}\,
\lambda^{(\psi)}_{CB},}\\[4pt]
{s\tpsi^{\alpha}_{iBf}=-i\,\tlambda^{(\phi)}_{ij}\,\tpsi^{\alpha}_{jBf}\,-\,i\,
\tpsi^{\alpha}_{iCf}\,\lambda^{(\psi)}_{CB},}\\[4pt]
{s\psi_{\alpha Aif'}=-i\,\tlambda^{(\psi)}_{AC}\,\psi_{\alpha Cif'}\,-\,i\,
\psi_{\alpha Ajf'}\lambda^{(\phi)}_{ji}.}\\
\end{array}
\label{nonirrbrs}
\end{equation}
In our notation, $\tlambda^{(\phi)}_{ij}=\lambda^{(\phi)}_{ji}$. The BRS
transformations in the previous eq. are a by-product of the BRS transformations
in eqs.~(\ref{brsphi}),~(\ref{brstipsi}) and~(\ref{brspsi}) and of  $\msC_{AiB}$
being, as shown in eq.~(\ref{invareq}), a group invariant tensor.

It can be seen~\cite{Aschieri:2002mc} that the naive noncommutative version of
$ \calY^{\text{(ord)}}$ as defined in eq.~(\ref{yukord})
 would not do, since, on the one hand, the $\star$-product is noncommutative
 and, on the other hand, the fact that the noncommutative gauge
transformations are valued on the universal enveloping algebra of the Lie
algebra  yields the conclusion that eq.~(\ref{invareq}) only leads to gauge
invariance at zero order in the noncommutative parameter.  By the naive noncommutative version of $\calY^{\text{(ord)}}$, we
mean the expression
\begin{equation*}
\idx\;\funY_{ff'}\;\msC_{AiB}\;\tPsi^{\alpha}_{A
  f}\;\star\;\Psi_{\alpha B f'}\;\star\;
\Phi_{i},
\end{equation*}
where $\tPsi^{\alpha}_{Af}$, $\Psi_{\alpha B f'}$ and $\Phi_{i}$ are defined
in terms of the ordinary fields by means of the standard --see eq. (3.3) in
ref.~\cite{Jurco:2001rq}-- Seiberg-Witten maps. However, if we include in our
formalism the notion of hybrid Seiberg-Witten
map introduced in ref.~\cite{Schupp:2001we}, one can naturally associate a
noncommutative Yukawa term to each $\calY_n^{\text{(ord)}}$, $n=1,2,3$, in
eq.~(\ref{yuk123}).  We shall see that the three noncommutative Yukawa terms
so obtained are not equal to one another, so our most general noncommutative
Yukawa term  will be the sum of them all.

To obtain the noncommutative version of $\calY_1^{\text{(ord)}}$ in eq.~(\ref{yuk123}),
one first introduces three noncommutative fields,
 $\tPsi^{\alpha}_{ A  f}$, $\Phi_{AB}$ and $\Psi_{\alpha B f'}$, which are, respectively,  the noncommutative counterparts of the ordinary fields,
 $\tpsi^{\alpha}_{ A  f}$, $\phi_{AB}$ and $\psi_{\alpha B f'}$
 in $\calY_1^{\text{(ord)}}$. The noncommutative fields are functions of the ordinary fields and $\theta^{\mu\nu}$ that solve the Seiberg-Witten map equations and go to its ordinary counterpart as $\theta^{\mu\nu}\rightarrow 0$. To define the Seiberg-Witten  map equations, one first introduces the noncommutative BRS transformations of $\tPsi^{\alpha}_{ A  f}$, $\Phi_{AB}$ and $\Psi_{\alpha B f'}$:
 \begin{equation}
 \begin{array}{l}
 {s_{\text{nc}}\tPsi^{\alpha}_{ A  f}=i\,\tPsi^{\alpha}_{B f}\star \tLambda^{(\psi)}_{BA},\quad s_{\text{nc}}\Psi_{\alpha B f'}=i\,\Lambda^{(\psi)}_{BC}\star\Psi_{\alpha C f'},}\\[4pt]
 {s_{\text{nc}}\Phi_{AB}=-i\,\tLambda^{(\psi)}_{AC}\star\Phi_{CB}-i\,\Phi_{AC}\star
\Lambda^{(\psi)}_{CB},}\\[4pt]
{s_{\text{nc}}\tLambda^{(\psi)}_{BA}=-i\,\tLambda^{(\psi)}_{BC}\star\tLambda^{(\psi)}_{CA},
\quad s_{\text{nc}}\Lambda^{(\psi)}_{BC}=i\,\Lambda^{(\psi )}_{BD}\star\Lambda^{(\psi)}_{DC}.}\\
\end{array}
\label{ncBRS}
\end{equation}
Let us stress that we have defined the noncommutative BRS transformation of $\tPsi^{\alpha}_{ A  f}$ by acting, via the $\star$ product, with $\tLambda^{(\psi)}_{BA}$ on the right of $\tPsi^{\alpha}_{ A  f}$. Hence, by definition, the noncommutative gauge transformations act on $\tPsi^{\alpha}_{ A  f}$ on the right. We shall see below that this right action makes the noncommutative Yukawa term gauge invariant, and it is to be compared with the noncommutative BRS  transformation of $\Psi_{\alpha B f}$ which is defined by
left action with the $\star$-product.

The Seiberg-Witten map eqs., which give
\begin{equation*}
\begin{array}{l}
{ \tPsi^{\alpha}_{ A f}[\widetilde{a}^{(\psi)}_\mu,\tpsi^{\alpha}_{Bf},\theta^{\mu\nu}],\quad \Phi_{AB}[\widetilde{a}^{(\psi)}_\mu,a^{(\psi)}_\mu,\phi_{AB},\theta^{\mu\nu}],\quad \Psi_{\alpha B f'}[a^{(\psi)}_\mu,\psi^{\alpha}_{\alpha Cf'},\theta^{\mu\nu}],}\\ {\tLambda^{(\psi)}_{BA}[\widetilde{a}^{(\psi)}_\mu,\tlambda^{(\psi)},\theta^{\mu\nu}]\quad \text{and}\quad \Lambda^{(\psi )}_{BC}[a^{(\psi)}_\mu,\lambda^{(\psi)},\theta^{\mu\nu}]}
\end{array}
\end{equation*}
as a function of their arguments, are the following:
\begin{equation}
\begin{array}{l}
{s_{\text{nc}}\tLambda^{(\psi)}_{BA}=s\tLambda^{(\psi)}_{BA},\quad s_{\text{nc}}\Lambda^{(\psi)}_{BA}=
s\Lambda^{(\psi)}_{BA},}\\[4pt]
{s_{\text{nc}}\tPsi^{\alpha}_{ A  f}=s\tPsi^{\alpha}_{ A  f},\quad
s_{\text{nc}}\Psi_{\alpha B f'}=s\Psi_{\alpha B f'},\quad
s_{\text{nc}}\Phi_{AB}=s\Phi_{AB}.}
\end{array}
\label{seiwiteqs}
\end{equation}
The symbol $s$ denotes the ordinary BRS operator defined in  eqs.~(\ref{brspsi}),~(\ref{brsphi}),~(\ref{brstipsi}) and~(\ref{nonirrbrs}), along with
\begin{equation}
\begin{array}{l}
{s\widetilde{a}^{(\psi)}_{\mu\,AB}=\partial_{\mu}\tlambda^{(\psi)}_{AB}
+i[\widetilde{a}^{(\psi)}_\mu,\tlambda^{(\psi)}]_{AB},\quad \widetilde{a}^{(\psi)}_{\mu\,AB}=a^a_{\mu}\tSigma^a_{AB},}\\[4pt]
{sa^{(\psi)}_{\mu\,AB}=\partial_{\mu}\lambda^{(\psi)}_{AB}
-i[a^{(\psi)}_\mu,\lambda^{(\psi)}]_{AB},\quad
a^{(\psi)}_{\mu\,AB}=a^a_{\mu}\Sigma^a_{AB}.}
\end{array}
\label{sapsi}
\end{equation}
Recall that $\tSigma^a_{AB}=\Sigma^a_{BA}$.

  Solutions to the Seiberg-Witten map eqs. in eq.~(\ref{seiwiteqs}) can be obtained as formal
powers series in $\theta^{\mu\nu}$. Up to first order, these solutions, which define the corresponding Seiberg-Witten maps, read
\begin{equation}
\begin{array}{l}
{\tLambda^{(\psi)}_{BA}=\tlambda^{(\psi)}_{BA}+\frac{1}{4}\,\theta^{\mu\nu}\,\{
\widetilde{a}^{(\psi)}_{\mu},\partial_{\nu}\tlambda^{(\psi)}\}_{BA}+O(\theta^2),}\\[4pt]
{\Lambda^{(\psi)}_{BC}=\lambda^{(\psi)}_{BC}-\frac{1}{4}\,\theta^{\mu\nu}\,\{
a^{(\psi)}_\mu,\partial_{\nu}\lambda^{(\psi)}\}_{BC}+O(\theta^2),}\\[4pt]
{\tPsi^{\alpha}_{ A f}=\tpsi^{\alpha}_{ A f}-\frac{1}{2}\,\theta^{\mu\nu}\,
\partial_{\mu}\tpsi^{\alpha}_{ B f}\widetilde{a}^{(\psi)}_{\nu\,BA}+
\frac{i}{4}\,\theta^{\mu\nu}\,\tpsi^{\alpha}_{ C f}\widetilde{a}^{(\psi)}_{\mu\,CB}
\widetilde{a}^{(\psi)}_{\nu\,BA}+O(\theta^2),}\\[4pt]
{\Phi_{ A B}=\phi_{ A B}+\frac{1}{2}\,\theta^{\mu\nu}\,
\widetilde{a}^{(\psi)}_{\mu\,AC}\partial_{\nu}\phi_{CB}+
\frac{i}{4}\,\theta^{\mu\nu}\,\widetilde{a}^{(\psi)}_{\mu\,AC}
\widetilde{a}^{(\psi)}_{\nu\,CD}\phi_{DB}+}\\[4pt]
{\phantom{\Phi_{AB}=\phi_{AB}}+\frac{1}{2}\,\theta^{\mu\nu}\,
\partial_{\mu}\phi_{AC}a^{(\psi)}_{\nu\,CB}+
\frac{i}{4}\,\theta^{\mu\nu}\,\phi_{AC}a^{(\psi)}_{\mu\,CD}
a^{(\psi)}_{\nu\,DB}}\\[4pt]
{\phantom{\Phi_{AB}=\phi_{AB}}+\frac{i}{2}\,\theta^{\mu\nu}\,
\widetilde{a}^{(\psi)}_{\mu\,AC}\phi_{CD}a^{(\psi)}_{\nu\,DB}+O(\theta^2),}\\[4pt]
{\Psi_{\alpha B f'}=\psi_{\alpha B f'}-\frac{1}{2}\,\theta^{\mu\nu}\,
a^{(\psi)}_{\mu\,BC}\partial_{\mu}\psi_{\alpha C f'}+
\frac{i}{4}\,\theta^{\mu\nu}\, a^{(\psi)}_{\mu\,BC}a^{(\psi)}_{\nu\,CD}\psi^{\alpha}_{ D f'}+O(\theta^2).}\\
\label{expanSWmap}
\end{array}
\end{equation}
Notice that $\Phi_{AB}$ is defined by a hybrid Seiberg-Witten map, a notion which was put forward in ref.~\cite{Schupp:2001we}.

We are now in the position to introduce and --using eq.~(\ref{expanSWmap})-- compute up to first order in $\theta^{\mu\nu}$ the noncommutative counterpart, $\calY_1^{\text{(nc)}}$, of $\calY_1^{\text{(ord)}}$
in eq.~(\ref{yuk123}):
\begin{equation}
\begin{array}{l}
{\calY_1^{\text{(nc)}}=
\idx\;\funY^{(1)}_{ff'}\;\tPsi^{\alpha}_{ A  f}\star\Phi_{AB}\star\Psi_{\alpha B f'}}\\[4pt]
{\phantom{\calY_1^{\text{(nc)}}}=\phantom{+}
\idx\;\funY^{(1)}_{ff'}\;\msC_{AiB}\;\tpsi^{\alpha}_{ A  f}\;\phi_i\;\psi_{\alpha B f'}}\\[4pt]
{\phantom{\calY_1^{\text{(nc)}}=}+
\idx\;(-\frac{i}{2})\,\theta^{\mu\nu}\,\funY^{(1)}_{ff'}\;\msC_{AiB}\;(D_{\mu}\tpsi^{\alpha}_{  f})_A\;\phi_i\;(D_{\nu}\psi_{\alpha f'})_B}\\[4pt]
{\phantom{\calY_1^{\text{(nc)}}=}+\idx\;(-\frac{1}{4})\,\big(\funY^{(1)}_{ff'}\;\msC_{AiB}-\funY^{(1)}_{f'f}\;\msC_{BiA}\big)
\;\theta^{\mu\nu}\;\phi_i\,\tpsi^{\alpha}_{ A  f}\;f^{(\psi)}_{\mu\nu\,BC}\;\psi_{\alpha C f'}+O(\theta^2),}\\
\end{array}
\label{ncy1}
\end{equation}
where $(D_{\mu}\tpsi^{\alpha}_{  f})_A=\partial_{\mu}\tpsi^{\alpha}_{A f}-i\tpsi^{\alpha}_{B f}\widetilde{a}^{(\psi)}_{\mu\,BA}$, $(D_{\nu}\psi_{\alpha f'})_B=\partial_{\nu}\psi_{\alpha B f'}-ia^{(\psi)}_{\nu\,BC}\psi_{\alpha C f'}$ and
$f^{(\psi)}_{\mu\nu}=\partial_{\mu}a^{(\psi)}_{\nu}-\partial_{\nu}a^{(\psi)}_{\mu}-
i[a^{(\psi)}_{\mu},a^{(\psi)}_{\nu}]$. It is apparent that $\calY_1^{\text{(nc)}}$ is invariant under the noncommutative BRS variations defined in eq.~(\ref{ncBRS}).
Next, we define the noncommutative counterpart, $\calY_2^{\text{(nc)}}$, of $\calY_2^{\text{(ord)}}$ in eq.~(\ref{yuk123}):
\begin{equation}
\calY_2^{\text{(nc)}}=
\idx\;\funY^{(2)}_{ff'}\;\tPhi_{i}\star\tPsi^{\alpha}_{ iB  f}\star\Psi_{\alpha B f'},
\label{ncy2}
\end{equation}
where
\begin{equation}
\begin{array}{l}
{\tPhi_i=\tphi_i-\frac{1}{2}\,\theta^{\mu\nu}\,
\partial_{\mu}\tphi_{j}\widetilde{a}^{(\phi)}_{\nu\,ji}+
\frac{i}{4}\,\theta^{\mu\nu}\,\tphi_{j}\widetilde{a}^{(\phi)}_{\mu\,jk}
\widetilde{a}^{(\phi)}_{\nu\,ki}+O(\theta^2),}\\[4pt]
{\tPsi^{\alpha}_{iBf}=\tpsi^{\alpha}_{i B f}+\frac{1}{2}\,\theta^{\mu\nu}\,
\widetilde{a}^{(\phi)}_{\mu\,ij}\partial_{\nu}\tpsi^{\alpha}_{jBf}+
\frac{i}{4}\,\theta^{\mu\nu}\,\widetilde{a}^{(\phi)}_{\mu\,ik}
\widetilde{a}^{(\phi)}_{\nu\,kj}\tpsi^{\alpha}_{jBf}+}\\[4pt]
{\phantom{\Phi_{AB}=\phi_{AB}}+\frac{1}{2}\,\theta^{\mu\nu}\,
\partial_{\mu}\tpsi^{\alpha}_{iCf}a^{(\psi)}_{\nu\,CB}+
\frac{i}{4}\,\theta^{\mu\nu}\,\tpsi^{\alpha}_{iDf}a^{(\psi)}_{\mu\,DC}
a^{(\psi)}_{\nu\,CB}}\\[4pt]
{\phantom{\Phi_{AB}=\phi_{AB}}+\frac{i}{2}\,\theta^{\mu\nu}\,
\widetilde{a}^{(\phi)}_{\mu\,ij}\tpsi^{\alpha}_{jCf}a^{(\psi)}_{\nu\,CB}+O(\theta^2),}\\[4pt]
{\Psi_{\alpha B f'}=\psi_{\alpha A f}-\frac{1}{2}\,\theta^{\mu\nu}\,
a^{(\psi)}_{\mu\,BC}\partial_{\mu}\psi_{\alpha C f'}+
\frac{i}{4}\,\theta^{\mu\nu}\, a^{(\psi)}_{\mu\,BC}a^{(\psi)}_{\nu\,CD}\psi^{\alpha}_{ D f'}+O(\theta^2),}\\
\label{expanSWmap2}
\end{array}
\end{equation}
with $\widetilde{a}^{(\phi)}_{\mu\,ij}=a^a_{\mu}\widetilde{M}^a_{ij}$, $\widetilde{M}^a_{ij}=M^a_{ji}$.
The noncommutative fields in the previous eq. are solutions to the following Seiberg-Witten map eqs.:
\begin{equation}
\begin{array}{l}
{-i\,\tLambda^{(\phi)}_{ij}\star \tPsi^{\alpha}_{jBf}-i\,\tPsi^{\alpha}_{iCf}\star
\Lambda^{(\psi)}_{CB}\equiv s_{\text{nc}}\tPsi^{\alpha}_{iBf}=s \tPsi^{\alpha}_{iBf},}\\[4pt]
{\phantom{+}i\,\Lambda^{(\psi)}_{BC}\star\Psi_{\alpha Cf'}\equiv s_{\text{nc}}\Psi_{\alpha Bf'}=s \Psi_{\alpha Bf'},}\\[4pt]
{\phantom{+}i\,\tPhi_{j}\star\tLambda^{(\phi)}_{ji}\equiv s_{\text{nc}}\tPhi_{i}=s\tPhi_{i},}\\[4pt]
{\phantom{+}i\,\Lambda^{(\psi)}_{AC}\star \Lambda^{(\psi)}_{CB}\equiv s_{\text{nc}}\Lambda^{(\psi)}_{AC}=s \Lambda^{(\psi)}_{AC},}\\[4pt]
{-i\,\tLambda^{(\phi)}_{ik}\star \tLambda^{(\phi)}_{kj}\equiv s_{\text{nc}}\tLambda^{(\phi)}_{ij}=s \tLambda^{(\phi)}_{ij},}
\end{array}
\label{swmapeq2}
\end{equation}
where
\begin{equation*}
\begin{array}{l}
{\tLambda^{(\phi)}_{ij}=\tlambda^{(\phi)}_{ij}+\frac{1}{4}\,\theta^{\mu\nu}\,\{
\widetilde{a}^{(\phi)}_{\mu},\partial_{\nu}\tlambda^{(\phi)}\}_{ij}+O(\theta^2),}\\[4pt]
{\Lambda^{(\psi)}_{BC}=\lambda^{(\psi)}_{BC}-\frac{1}{4}\,\theta^{\mu\nu}\,\{
a^{(\psi)}_\mu,\partial_{\nu}\lambda^{(\psi)}\}_{BC}+O(\theta^2),}
\end{array}
\end{equation*}
with $\tlambda^{(\phi)}_{ij}=\tlambda^a \widetilde{M}^a_{ij}$.  To check that the
Seiberg-Witten maps in eq.~(\ref{expanSWmap2}) are solutions to eq.~(\ref{swmapeq2}), one needs the following results:
\begin{equation}
s\widetilde{a}^{(\phi)}_{\mu\,ij}=\partial_{\mu}\tlambda^{(\phi)}_{ij}
+i[\widetilde{a}^{(\phi)}_\mu,\tlambda^{(\phi)}]_{ij},\quad
sa^{(\phi)}_{\mu\,ij}=\partial_{\mu}\lambda^{(\phi)}_{ij}
-i[a^{(\phi)}_\mu,\lambda^{(\phi)}]_{ij},
\label{saphi}
\end{equation}
where $a^{(\phi)}_{\mu\,ij}=a^{a}_{\mu\,ij}\,M^{a}_{ij}$.

 By using the results in eq.~(\ref{expanSWmap2}), one obtains the  $\theta$-expansion of $\calY_2^{\text{(nc)}}$ in eq.~(\ref{ncy2}):
\begin{equation}
\begin{array}{l}
{\calY_2^{\text{(nc)}}=\phantom{+}
\idx\;\funY^{(2)}_{ff'}\;\msC_{AiB}\;\tpsi^{\alpha}_{ A  f}\;\phi_i\;\psi_{\alpha B f'}}\\[4pt]
{\phantom{\calY_1^{\text{(nc)}}=}+
\idx\;(\frac{i}{2})\,\theta^{\mu\nu}\,\funY^{(2)}_{ff'}\;\msC_{AiB}\;(D_{\mu}\tpsi^{\alpha}_{  f})_A\;\phi_i\;(D_{\nu}\psi_{\alpha f'})_B}\\[4pt]
{\phantom{\calY_1^{\text{(nc)}}=}+\idx\;(-\frac{1}{4})\,\big(\funY^{(2)}_{ff'}\;\msC_{AiB}+\funY^{(2)}_{f'f}\;\msC_{BiA}\big)
\;\theta^{\mu\nu}\;\phi_i\,\tpsi^{\alpha}_{ A  f}\;f^{(\psi)}_{\mu\nu\,BC}\;\psi_{\alpha C f'}+O(\theta^2).}\\
\end{array}
\label{expancy2}
\end{equation}
In obtaining the previous result, the following eq. is of much help:
\begin{equation}
\widetilde{f}^{(\psi)}_{\mu\nu\, AC}\,\msC_{CiB}\,+\,\msC_{AjB}
f^{(\phi)}_{\mu\nu\,ji}\,+\,\msC_{AiC}\,f^{(\psi)}_{\mu\nu\,CB}=0.
\label{fmunueq}
\end{equation}
Notice that $\widetilde{f}^{(\psi)}_{\mu\nu}=\partial_{\mu}\widetilde{a}^{(\psi)}_{\nu}-
\partial_{\nu}\widetilde{a}^{(\psi)}_{\mu}+
i[\widetilde{a}^{(\psi)}_{\mu},\widetilde{a}^{(\psi)}_{\nu}]$ and
$f^{(\phi)}_{\mu\nu}=\partial_{\mu}a^{(\phi)}_{\nu}-
\partial_{\nu}a^{(\phi)}_{\mu}-
i[a^{(\phi)}_{\mu},a^{(\phi)}_{\nu}]$.
Eq.~(\ref{fmunueq}), and  similar eqs. involving $a^{(\psi)}_{\mu}$ and $a^{(\phi)}_{\mu}$, follow from eq.~(\ref{invareq}).

Finally, we shall introduce the noncommutative version, $\calY_3^{\text{(nc)}}$, of $\calY_3^{\text{(ord)}}$ in eq.~(\ref{yuk123}):
\begin{equation}
\calY_3^{\text{(nc)}}=
\idx\;\funY^{(3)}_{ff'}\;\tPsi^{\alpha}_{ A  f}\star\Psi_{\alpha A i f'}\star \Phi_{i}.
\label{ncy3}
\end{equation}
The fields in the previous eq. are given, at first order in $\theta$, by the following expressions:
\begin{equation}
\begin{array}{l}
{\tPsi^{\alpha}_{ A f}=\tpsi^{\alpha}_{ A f}-\frac{1}{2}\,\theta^{\mu\nu}\,
\partial_{\mu}\tpsi^{\alpha}_{ B f}\widetilde{a}^{(\psi)}_{\nu\,BA}+
\frac{i}{4}\,\theta^{\mu\nu}\,\tpsi^{\alpha}_{ C f}\widetilde{a}^{(\psi)}_{\mu\,CB}
\widetilde{a}^{(\psi)}_{\nu\,BA}+O(\theta^2),}\\[4pt]
{\Psi_{\alpha Aif'}=\psi_{\alpha Aif'}+\frac{1}{2}\,\theta^{\mu\nu}\,
\widetilde{a}^{(\psi)}_{\mu\,AB}\partial_{\nu}\psi_{\alpha Bif'}+\frac{i}{4}
\widetilde{a}^{(\psi)}_{\mu\,AB}\widetilde{a}^{(\psi)}_{\nu\,BC}\psi_{\alpha Cif'}}\\[4pt]
{\phantom{\Psi_{\alpha Aif'}=\psi_{\alpha Aif'}}
+\frac{1}{2}\,\theta^{\mu\nu}\,\partial_{\mu}\psi_{\alpha Ajf'}a^{(\phi)}_{\nu\,ji}+\frac{i}{4}\psi_{\alpha Akf'}a^{(\phi)}_{\mu\,kj}a^{(\phi)}_{\nu\,ji}}\\[4pt]
{\phantom{\Psi_{\alpha Aif'}=\psi_{\alpha Aif'}}
+\frac{i}{2}\,\theta^{\mu\nu}\,\widetilde{a}^{(\psi)}_{\mu\,AB}\psi_{\alpha Bjf'}a^{(\phi)}_{\nu\,ji}+O(\theta^2),}\\[4pt]
{\Phi_{i}=\phi_{i}-\frac{1}{2}\,\theta^{\mu\nu}\,a^{(\phi)}_{\mu\,ij}\partial_{\nu}\phi_{j}+
\frac{i}{4}\,\theta^{\mu\nu}\,a^{(\phi)}_{\mu\,ij}a^{(\phi)}_{\nu\,jk}\phi_{k}
+O(\theta^2).}
\label{expanSWmap3}
\end{array}
\end{equation}
The Seiberg-Witten maps in the previous set of eqs. are solutions to
\begin{equation}
\begin{array}{l}
{\phantom{-}i\,\tPsi^{\alpha}_{Bf}\star \tLambda^{(\psi)}_{BA}\equiv s_{\text{nc}}\tPsi^{\alpha}_{Af}=s\tPsi^{\alpha}_{Af},}\\[4pt]
{-i\,\tLambda^{(\psi)}_{AC}\star \Psi_{\alpha Cif'}-i\,\Psi_{\alpha Ajf'}\star\Lambda^{(\phi)}_{ji}\equiv s_{\text{nc}}\Psi_{\alpha Aif'}=s\Psi_{\alpha Aif'},}\\[4pt]
{\phantom{-}i\,\Lambda^{(\phi)}_{ij}\star \Phi_j\equiv s_{\text{nc}}\Phi_i=s\Phi_i ,}\\[4pt]
{-i\,\tLambda^{(\psi)}_{AC}\star \tLambda^{(\psi)}_{CB}\equiv s_{\text{nc}}
\tLambda^{(\psi)}_{AB}=s\tLambda^{(\psi)}_{AB},}\\[4pt]
{i\,\Lambda^{(\phi)}_{ik}\star \Lambda^{(\phi)}_{kj}\equiv s_{\text{nc}}
\Lambda^{(\phi)}_{ij}=s\Lambda^{(\phi)}_{ij},}\\[4pt]
\end{array}
\label{swmapeq3}
\end{equation}
if
\begin{equation*}
\begin{array}{l}
{\Lambda^{(\phi)}_{ij}=\lambda^{(\phi)}_{ij}-\frac{1}{4}\,\theta^{\mu\nu}\,\{
a^{(\phi)}_{\mu},\partial_{\nu}\lambda^{(\phi)}\}_{ij}+O(\theta^2),}\\[4pt]
{\tLambda^{(\psi)}_{AB}=\tlambda^{(\psi)}_{AB}+\frac{1}{4}\,\theta^{\mu\nu}\,\{
\widetilde{a}^{(\psi)}_\mu,\partial_{\nu}\tlambda^{(\psi)}\}_{AB}+O(\theta^2).}
\end{array}
\end{equation*}

Now, substituting the Seiberg-Witten maps in eq.~(\ref{expanSWmap3}) in eq.~(\ref{ncy3}), one gets
\begin{equation}
\begin{array}{l}
{\calY_3^{\text{(nc)}}=\phantom{+}
\idx\;\funY^{(3)}_{ff'}\;\msC_{AiB}\;\tpsi^{\alpha}_{ A  f}\;\phi_i\;\psi_{\alpha B f'}}\\[4pt]
{\phantom{\calY_1^{\text{(nc)}}=}+
\idx\;(\frac{i}{2})\,\theta^{\mu\nu}\,\funY^{(3)}_{ff'}\;\msC_{AiB}\;(D_{\mu}\tpsi^{\alpha}_{  f})_A\;\phi_i\;(D_{\nu}\psi_{\alpha f'})_B}\\[4pt]
{\phantom{\calY_1^{\text{(nc)}}=}+\idx\;(\frac{1}{4})\,\big(\funY^{(3)}_{ff'}\;\msC_{AiB}+\funY^{(3)}_{f'f}\;\msC_{BiA}\big)
\;\theta^{\mu\nu}\;\phi_i\,\tpsi^{\alpha}_{ A  f}\;f^{(\psi)}_{\mu\nu\,BC}\;\psi_{\alpha C f'}+O(\theta^2).}\\
\end{array}
\label{expancy3}
\end{equation}

We have found no reason to discard any of the $\calY_n^{\text{(nc)}}$, $n=1,2,3$, in eqs.~(\ref{ncy1}), ~(\ref{ncy2}) and ~(\ref{ncy3}), respectively,
as a valid noncommutative Yukawa contribution, we then conclude that our
noncommutative Yukawa term, $\calY^{\text{(nc)}}$, is the sum of the three of them:
\begin{equation}
\calY^{\text{(nc)}}\equiv \calY_1^{\text{(nc)}}+\calY_2^{\text{(nc)}}+
\calY_3^{\text{(nc)}}.
\label{finncy}
\end{equation}
Using the expansions in eqs.~(\ref{ncy1}), ~(\ref{expancy2}) and ~(\ref{expancy3}), one can show that the most general functional which is linear in $\theta^{\mu\nu}$, contains  one $\phi_i$ and  two $\psi_{\alpha Af}$, involves the derivatives of
these fields, has
no dimensionful parameter other than $\theta^{\mu\nu}$  and whose BRS  variation vanishes, is given by the first order in $\theta$ contribution to $\calY^{\text{(nc)}}$ above. Hence, the noncommutative
Yukawa interaction introduced in  eq.~(\ref{finncy}) is renormalisable at first  order in $\theta^{\mu\nu}$: a property not to be overlooked.

\section{Taking into account the index symmetry properties of $\msC_{AiB}$}

Let $\phi_i$ in eq.~(\ref{yukord}) carry an irreducible representation of
SO(10), and, let $\msC_{AiB}$ be the invariant tensor also in
eq.~(\ref{yukord}). Then, the Clebsch-Gordan decomposition~\cite{Frappat} of the $16\bigotimes 16$
representation of SO(10)  leads to the conclusion that
$\msC_{AiB}=\msC_{BiA}$, if $\phi_i$ carries either the 10 or the
$\overline{126}$ of SO(10), and, that $\msC_{AiB}=-\msC_{BiA}$, if $\Phi_i$ transforms
under the 120 of SO(10). Analogously~\cite{Frappat}, that, for $\text{E}_6$, we have $27\bigotimes
27=(\overline{27}\bigoplus \overline{351'})_{\text{s}}\bigoplus \overline{351}_{\text{as}}$, implies that
$\msC_{AiB}=\msC_{BiA}$, when the Higgs field is in either the 27 or the
$351'$ of $\text{E}_6$, and $\msC_{AiB}=-\msC_{BiA}$, when $\phi_i$
carries the $351$ of $\text{E}_6$.

 That in our case $\msC_{AiB}$ has well-defined symmetry properties under
the exchange of $``A"$ and $``B"$ leads to simplified expressions for
$\calY^{\text{(nc)}}$ in  eq.~(\ref{finncy}). Indeed, if
$\msC_{AiB}=\msC_{BiA}$, eqs.~(\ref{ncy1}),~(\ref{expancy2}),~(\ref{expancy3})
and~(\ref{finncy}) yield
\begin{equation*}
\begin{array}{l}
{\calY^{\text{(nc)}}=\phantom{+}
\idx\;\funY^{(\text{s})}_{ff'}\;\msC_{AiB}\;\tpsi^{\alpha}_{ A  f}\;\phi_i\;\psi_{\alpha B f'}}\\[4pt]
{\phantom{\calY^{\text{(nc)}}=}+\idx\;(\frac{i}{2})\,\big(-\funY^{(1,\text{as})}_{ff'}+\funY^{(2,\text{as})}_{ff'}+\funY^{(3,\text{as})}_{ff'}\big)
\,\theta^{\mu\nu}\,\msC_{AiB}\,(D_{\mu}\tpsi^{\alpha}_{  f})_A\;\phi_i\;(D_{\nu}\psi_{\alpha f'})_B}\\[4pt]
{\phantom{\calY_1^{\text{(nc)}}=}+\idx\;(-\frac{1}{2})\,\big(\funY^{(1,\text{as})}_{ff'}+\funY^{(2,\text{s})}_{ff'}-\funY^{(3,\text{s})}_{ff'}\big)\,\theta^{\mu\nu}\,
\msC_{AiB}\,\phi_i\,\tpsi^{\alpha}_{ A  f}\;f^{(\psi)}_{\mu\nu\,BC}\;\psi_{\alpha C f'}+O(\theta^2),}\\
\end{array}
%\label{expancyncsym}
\end{equation*}
where
$\funY^{(\text{s})}_{ff'}=\funY^{(1,\text{s})}_{ff'}+\funY^{(2,\text{s})}_{ff'}+\funY^{(3,\text{s})}_{ff'}$.
$\funY^{(n,\text{s})}_{ff'}$ and $\funY^{(n,\text{as})}_{ff'}$ denote, respectively, the
symmetric and antisymmetric parts of $\funY^{(n)}_{ff'}$, with regard to the
indices $f,f'$. $\funY^{(n)}_{ff'}$, $n=1,2,3$ were introduced in
eqs.~(\ref{ncy1}),~(\ref{ncy2}) and~(\ref{ncy3}). Similarly, when
$\msC_{AiB}=-\msC_{BiA}$, eq.~(\ref{finncy}) boils down to
\begin{equation*}
\begin{array}{l}
{\calY^{\text{(nc)}}=\phantom{+}
\idx\;\funY^{(\text{as})}_{ff'}\;\msC_{AiB}\;\tpsi^{\alpha}_{ A  f}\;\phi_i\;\psi_{\alpha B f'}}\\[4pt]
{\phantom{\calY^{\text{(nc)}}=}+\idx\;(\frac{i}{2})\,\big(-\funY^{(1,\text{s})}_{ff'}+\funY^{(2,\text{s})}_{ff'}+\funY^{(3,\text{s})}_{ff'}\big)
\,\theta^{\mu\nu}\,\msC_{AiB}\,(D_{\mu}\tpsi^{\alpha}_{  f})_A\;\phi_i\;(D_{\nu}\psi_{\alpha f'})_B}\\[4pt]
{\phantom{\calY_1^{\text{(nc)}}=}+\idx\;(-\frac{1}{2})\,\big(\funY^{(1,\text{s})}_{ff'}+\funY^{(2,\text{as})}_{ff'}-\funY^{(3,\text{as})}_{ff'}\big)\,\theta^{\mu\nu}\,
\msC_{AiB}\,\phi_i\,\tpsi^{\alpha}_{ A  f}\;f^{(\psi)}_{\mu\nu\,BC}\;\psi_{\alpha C f'}+O(\theta^2),}\\
\end{array}
%\label{expancyncasym}
\end{equation*}
where $\funY^{(\text{as})}_{ff'}=\funY^{(1,\text{as})}_{ff'}+\funY^{(2,\text{as})}_{ff'}+\funY^{(3,\text{as})}_{ff'}$.

\section{Redundant choices}

Recall that $\tPsi^{\alpha}_{iBf}$ is the noncommutative counterpart of
$\tpsi^{\alpha}_{iBf}$ in eq.~(\ref{reducifields}).
The reader may rightly ask whether a new Yukawa term can be obtained by
making the following choice --to be compared with the definition in eq.~(\ref{swmapeq2})-- for the noncommutative BRS transformations of
$\tPsi^{\alpha}_{iBf}$:
\begin{equation}
s_{\text{nc}}\tPsi^{\alpha}_{iBf}=-i\, \tPsi^{\alpha}_{jBf}\star \tLambda^{(\phi)}_{ij}
-i\,\Lambda^{(\psi)}_{CB}\star  \tPsi^{\alpha}_{iCf}.
\label{opsnc}
\end{equation}
Notice that this is a noncommutative generalisation of the  BRS transformations, in eq.~(\ref{nonirrbrs}),   of   $\tpsi^{\alpha}_{iBf}$. Also notice that we  go back to $s_{\text{nc}}\tPsi^{\alpha}_{iBf}$ in eq.~(\ref{swmapeq2}),
when we change the order in which the $\Lambda$'s and $\tPsi^{\alpha}_{iBf}$ occur in eq.~(\ref{opsnc}).   Since the way in which the contracted indices occur in
eq.~(\ref{opsnc}) is a little odd, we shall rename the objects in that eq. as follows:
\begin{equation*}
\tPsi^{\alpha}_{iBf}\equiv\Psi^{'\,\alpha}_{iBf},\quad
\tLambda^{(\phi)}_{ij}\equiv\Lambda^{'\,(\phi)}_{ji},\quad
\Lambda^{(\psi)}_{CB}\equiv  \tLambda^{'\,(\psi)}_{BC}.
\end{equation*}
In terms of the fields we have just introduced eq.~(\ref{opsnc}) reads
\begin{equation}
s_{\text{nc}}\Psi^{'\,\alpha}_{Bif}=-i\, \tPsi^{'\,\alpha}_{Bjf}
\star \Lambda^{'\,(\phi)}_{ji}
-i\,\tLambda^{'\,(\psi)}_{BC}\star  \tPsi^{'\,\alpha}_{Cif}.                                                \label{opsncord}
\end{equation}
This eq. is to be supplemented with
\begin{equation}
s_{\text{nc}}\Lambda^{'\,(\phi)}_{ji}=i \Lambda^{'\,(\phi)}_{jk}
\star \Lambda^{'\,(\phi)}_{ki},\quad
s_{\text{nc}}\tLambda^{'\,(\psi)}_{BC}=-i\tLambda^{'\,(\psi)}_{BD}
\star \tLambda^{'\,(\psi)}_{DC},
\label{nilbrs}
\end{equation}
if we want $s_{\text{nc}}^{2}=0$.

Let us next  introduce $\Phi{'}_i$ and $\tPsi^{'}_{\alpha iBf'}$  as  the new noncommutative counterparts of the ordinary $\phi_i$ and
$\tpsi^{'}_{\alpha Bf'}=\psi^{'}_{\alpha Bf'}$, the latter entering  the
ordinary Yukawa term in eq.~(\ref{yukord}). The BRS transformations of
$\Phi{'}_i$ and $\tPsi^{'}_{\alpha iBf'}$ are defined as follows:
\begin{equation}
s_{\text{nc}}\tPsi^{'}_{\alpha Af}\equiv
i\,\tPsi^{'}_{\alpha Bf}\star \tLambda^{'}_{BA},\quad
s_{\text{nc}}\Phi^{'}_{i}\equiv
i\,\Lambda^{'\, (\phi)}_{ij}\star\Phi^{'}_{j}.
\label{newPhi}
\end{equation}
Now, it is plain that
\begin{equation}
\calY_4^{\text{(nc)}}=
\idx\;\funY^{(4)}_{f'f}\;\tPsi^{'\,\alpha}_{ A  f'}\star\Psi^{'}_{\alpha A i f}\star
\Phi^{'}_{i}
\label{ncy4}
\end{equation}
is invariant under noncommutative BRS transformations, if the fields in it are
solutions  to the following Seiberg-Witten map eqs.:
\begin{equation}
s_{\text{nc}}\tPsi^{'\,\alpha}_{ A  f'}=s\tPsi^{'\,\alpha}_{ A  f'},\quad
s_{\text{nc}}\Psi^{'}_{\alpha Bif}= s  \Psi^{'}_{\alpha Bif},\quad
 s_{\text{nc}}\Phi^{'}_{i}=s \Phi^{'}_{i},\quad
 s_{\text{nc}}\Lambda^{'\,(\phi)}_{ji}=s\Lambda^{'\,(\phi)}_{ji},\quad
 s_{\text{nc}}\tLambda^{'\,(\psi)}_{BC}=s\tLambda^{'\,(\psi)}_{BC},
 \label{swmapeq4}
\end{equation}
where the action of the noncommutative  BRS operator, $s_{\text{nc}}$, is defined
in eqs.~(\ref{opsncord}), (\ref{nilbrs}) and~(\ref{newPhi}), and the ordinary BRS operator, $s$, is given in eqs.~(\ref{brspsi}),~(\ref{brsphi}),~(\ref{brstipsi}),
~(\ref{nonirrbrs}),~(\ref{sapsi}) and~(\ref{saphi}). However, the Yukawa term in
eq.~(\ref{ncy4}) is not a new Yukawa term, but
it is the Yukawa term in eq.~(\ref{ncy3}). Indeed, notice that $\it{i})$  the   Seiberg-Witten  map equations in eq.~(\ref{swmapeq4}) are those in
eq.~(\ref{swmapeq3}) and $\it{ii})$ that at $\theta^{\mu\nu}=0$
the solutions to eq.~(\ref{swmapeq4}) must satisfy
\begin{equation*}
\begin{array}{l}
{\tPsi^{'\,\alpha}_{ A  f'}[\theta=0]= \tpsi^{\alpha}_{ A  f'},\quad
\Psi^{'}_{\alpha Bif}[\theta=0]=\tpsi_{\alpha iBf}\equiv
\tpsi_{\alpha Af}\,\msC_{AiB},\quad \Phi^{'}_{i}[\theta=0]=\phi_{i},}\\[4pt]
{\Lambda^{'\,(\phi)}_{ji}[\theta=0]=\lambda^{(\phi)}_{ji},\quad
\tLambda^{'\,(\psi)}_{BC}[\theta=0]= \tlambda^{(\psi)}_{BC}.}
\end{array}
\end{equation*}
Then, the fact that $\msC_{AiB}=\pm \msC_{BiA}$ --see previous section-- leads to
$\tpsi_{\alpha Af}\,\msC_{AiB}=\pm  \msC_{BiA}\, \psi_{\alpha Af}\equiv \pm\psi_{\alpha Bif}$, which combined with $\it{i})$ and $\it{ii})$ above   implies  that
\begin{equation}
\tPsi^{'\,\alpha}_{ A  f'}=\tPsi^{\alpha}_{ A  f'},\quad
\Psi^{'}_{\alpha Bif}=\pm \Psi_{\alpha Bif},\quad \Phi^{'}_{i}=\Phi_{i},
\label{idensol}
\end{equation}
where  $\tPsi^{\alpha}_{ A  f'}$, $\Psi_{\alpha Bif}$  and $\Phi_{i}$ are
the solutions to eq.~(\ref{swmapeq3}) whose first-order-in-$\theta$ expansions
are displayed in eq.~(\ref{expanSWmap3}). Finally, by substituting
eq.~(\ref{idensol}) in eq.~(\ref{ncy4}), one recovers eq.~(\ref{ncy3}). We thus
conclude that the Yukawa term in eq.~(\ref{ncy4}) is redundant.

Analogously, if  the  fields $\Psi_{\alpha Aif'}$ and $\Phi_{AB}$
--which are, respectively, the noncommutative counterparts of   the ordinary fields
$\psi_{\alpha Aif'}$ and $\phi_{AB}$ in eq.~(\ref{reducifields})-- are defined so that their noncommutative BRS transformations are given by
\begin{equation}
 s_{\text{nc}}\Psi_{\alpha Aif'}=-i\,\Psi_{\alpha Cif'}\star\tLambda^{(\psi)}_{AC}
 -i\,\Lambda^{(\phi)}_{ji}\star \Psi_{\alpha  Ajf'},\quad
 s_{\text{nc}}\Phi_{AB}=-i\,\Phi_{CB}\star \tLambda^{(\psi)}_{AC}-i\,
 \Lambda^{(\psi)}_{CB}\star \Phi_{AC},
 \label{newBRS}
\end{equation}
one may show that no new Yukawa terms arise out of them. Indeed, proceeding similarly as we did above, one may show that $\Psi_{\alpha Aif'}$ and $\Phi_{AB}$
transforming as in eq.~(\ref{newBRS}) yield $\calY_2^{\text{(nc)}}$ and
$\calY_1^{\text{(nc)}}$, respectively.
$\calY_2^{\text{(nc)}}$ is given in eq.~(\ref{ncy2}) and $\calY_1^{\text{(nc)}}$
was introduced in eq.~(\ref{ncy1}).

A last remark, the two $\Lambda$'s in the noncommutative BRS transformations of
$\Phi_{AB}$,  $\tPsi^{\alpha}_{iBf}$ and $\Psi_{\alpha Aif'}$ cannot both occur,
in the BRS transformation, on the same side of the corresponding field, for then,  $s_{\text{nc}}^2$ will not
vanish when acting on those fields, which in turn will render meaningless the Seiberg-Witten map eqs. for $\Phi_{AB}$,  $\tPsi^{\alpha}_{iBf}$ and $\Psi_{\alpha Aif'}$ --recall that $s^2=0$, if $s$ is the ordinary BRS operator.

\section{Conclusions}

We have seen in this paper that noncommutative Yukawa GUT terms can be
constructed in a natural way by applying the enveloping-algebra formalism
to ordinary fields --$\phi_{AB}$, $\tpsi^{\alpha}_{iBf}$ and $\psi_{\alpha
  Aif'}$ in eq.~(\ref{reducifields}), which transform under reducible
representations of the gauge group, but, which involve the very same number of
physical degrees as the ordinary irreducible multiplets they are made out of.
Let us stress that in the noncommutative case, in sharp contrast with
ordinary case, Yukawa terms cannot be constructed, in general --and, in
particular, for SO(10) and $\text{E}_6$-- by applying the Seiberg-Witten map
to ordinary irreducible multiplets, so, other procedures such as the one
put forward in this paper are needed. Our procedure, which takes advantage of
the notion of hybrid Seiberg-Witten map introduced in
ref.~\cite{Schupp:2001we},  yields a renormalisable
Yukawa term at first order in $\theta$, thus paving the way --in view of the
results in ref.~\cite{Martin:2009vg}-- to constructing
renormalisable noncommutative SO(1O) and$\text{E}_6$  GUTs; at least, at first
order in $\theta^{\mu\nu}$. Of course, the
next challenging issue is to define a noncommutative Higgs potential
which deforms the already involved --see, eg, refs.~\cite{Harvey:1981hk}  and~\cite{Babu:1984mz}--
ordinary GUT Higgs potential. This, although certainly
feasible within the noncommutative GUT formalism of ref.~\cite{Aschieri:2002mc}
 with  help from the ideas presented  in this paper, is a much involved piece of
 research and deserves a separate study.  Let us finally point out that
   eqs.~(\ref{ncy1}), ~(\ref{ncy2}) and ~(\ref{ncy3}) generalise naively to higher
   space-time dimensions, so the procedure introduced in this paper to construct
   Yukawa terms may be of help in formulating GUTs in higher dimensional
   noncommutative space-times~\cite{Aschieri:2004vh, Mondragon:2008dx, Cecotti:2009zf}.

\section{Acknowledgements}
I am indebted to Professor P. Schupp for very useful comments on the content of this paper.
This work has been financially supported in part by MICINN through grant
FPA2008-04906 and also in part by UCM-Banco Santander via grant GR58/08 910770.
%%%%%%%%%%%%%%%%%%%%%%%%%%%%%%%%%%%%%%%%%%%%%%%%%%%%%%%%
%%%%%%%%%%%%%%%%%%%%%%%%%%%%%%%%%%%%%%%%%%%%%%%%%%%%%%%%%
%%%%%%%%%%%%%%%%%%%%%%%%%%%%%%%%%%%%%%%%%%%%%%%%%%

\end{document}